\documentclass[12pt]{article}
\begin{document}

\begin{center}
{\Large \bf
Inequalities for the norms of vector functions

in a spherical layer

\vspace{5mm}

Valery V. Denisenko$^{1}$, Semen A. Nesterov$^{1}$
}
\end{center}

\vspace{2mm}

$^{1}$ {Institute of Computational Modelling, Russian Academy of Sciences, Siberian Branch

660036 Krasnoyarsk, Russia, denisen@icm.krasn.ru}

\vspace{2mm}

{\bf Abstract}

We consider the vector functions in a domain homeomorphic to a spherical layer bounded by twice continuously differentiable surfaces.
Additional restrictions are imposed on the domain, which allow to conduct proofs using simple methods.
On the outer and inner boundaries, the normal and the tangential components of the vector are zero, respectively.
For such functions, the integral over the domain of the squared vector
is estimated from above via the integral of the sum of squared gradients of its Cartesian components.
The last integral is estimated through the integral of the sum of the squared divergence and rotor.
These inequalities allow to define two norms equivalent to
the sum of the norms of the Cartesian components of vector functions as the elements of the space $W_2^{(1)}(\Omega)$.
The integrals over the boundaries of the squared vector are also estimated.
The constants in all proved inequalities are determined only by the shape of the domain and do not depend on a specific vector function.
The inequalities are necessary for investigating the operator of a mixed elliptic boundary value problem.

\vspace{3mm}
{\bf Key words:}
space of the vector functions, equivalent norms, mixed elliptic boundary value problem, spherical layer

\section*{Introduction}
We consider the set of smooth vector functions ${\bf P}$ in
the domain $\Omega$ that is bounded and homeomorphic to the spherical layer.
Its outer $\Gamma$ and inner $\gamma$ boundaries are twice continuously differentiable surfaces and homeomorphic to the sphere.
Normal (positive outward direction) and tangent to the boundary components of vectors
are marked with the indices $n$ and $\tau$.

The purpose of this work is to prove two inequalities for the vector function ${\bf P}$, satisfying the conditions:
\begin{equation}
P_n \mid_{\Gamma}=0, \quad \mathbf{P}_{\tau}\mid_{\gamma}=0.
\label{eq:11}
\end{equation}

The first inequality
\begin{eqnarray}
\int |{\bf P}|^2\,d\Omega\leq C_1\int |\mbox{grad}\,{\bf P}|^2\,d\Omega,
\label{eq:3}
\end{eqnarray}
where $|\mbox{grad}\,{\bf P}|^2$ is the sum of squared modulus of the gradients of all Cartesian components of ${\bf P}$,
and the constant $C_1$ is determined only by the shape of the domain $\Omega$ and does not depend on the specific function ${\bf P}$.

This inequality allows to use its right-hand side as the norm ${\bf P}$, which is equivalent to
the sum of the norms of the Cartesian components ${\bf P}$ as the elements of the space $W_2^{(1)}(\Omega)$.

This inequality is similar to the Friedrichs inequality for scalar functions which are equal to zero at the boundary.
If the entire vector ${\bf P}$ is equal to zero at the boundary, for each of its Cartesian components one can use
Friedrichs inequality to obtain the required inequality. However, of interest are the functions
with only normal or only tangent components equal to zero at the boundary.

The second required inequality
\begin{eqnarray}
\int ((\mbox{rot}\,{\bf P})^2+(\mbox{div}\,{\bf P})^2)\,d\Omega\geq
C_2\int |\mbox{grad}\,{\bf P}|^2\,d\Omega,
\label{eq:4}
\end{eqnarray}
where the constant $C_2$ is also determined only by the shape of the domain.

Similar inequalities are used in the study of the orthogonal decomposition
of the space of the vector functions \cite{Bykhovsky_Smirnov}
and allow investigating the properties of operators of the elliptic boundary value problems \cite{Denisenko_1997}.

It is easy to check the inequality
\begin{eqnarray}
(\mbox{rot}\,{\bf P})^2+(\mbox{div}\,{\bf P})^2\leq 3\,|\mbox{grad}\,{\bf P}|^2.
\nonumber
\end{eqnarray}

Together with (\ref{eq:3}, \ref{eq:4})
this inequality allows to use the left-hand side of (\ref{eq:4}) as one more norm of ${\bf P}$, that is equivalent to
the sum of the norms of the Cartesian component of ${\bf P}$ as the elements of the space $W_2^{(1)}(\Omega)$.

Both inequalities (\ref{eq:3}, \ref{eq:4}) for the functions with one of the conditions (\ref{eq:11}),
set on the entire boundary of an arbitrary multiply connected domain are proved in the paper \cite{Bykhovsky_Smirnov}.
In applied research, mixed boundary value problems sometimes arise when conditions at different sections of the boundary
differ, for example, have the form (\ref{eq:11}), and for this case new proofs are required.

We also get estimates of the traces of the vector function on the boundary of the domain.

\section{The domains}
To limit ourselves to simple means, we make additional assumptions:
both surfaces are smooth, the surface $\Gamma$ is convex, the curvature of $\gamma$ is bounded,
and we also consider the differences of both surfaces to be limited from two spheres with a common center and not too different radii.
Let's give these constraints a specific look.

We consider these surfaces to be defined in spherical coordinates using the functions $R_\Gamma(\theta,\varphi)$
and $R_\gamma(\theta,\varphi)$, such that
\begin{eqnarray}
0<R_1\leq R_\gamma(\theta,\varphi)\leq R_2, \quad 0<R_\Gamma(\theta,\varphi)-R_\gamma(\theta,\varphi)\leq \delta R,  \quad
\quad R_\Gamma(\theta,\varphi)\leq R_3.
\label{eq:5}
\end{eqnarray}

It is convenient to write down the convexity condition for the surface $\Gamma$ in local form.
At an arbitrary point of $\Gamma$, construct a tangent plane with Cartesian coordinates $ x, y $ and direct the $ z $ axis along the outward normal.
The surface equation can be written as
\begin{eqnarray}
z=\tilde f(x,y)
\label{eq:16}
\end{eqnarray}
with a twice continuously differentiable function $\tilde f(x,y)$ which is obtained by passing to local coordinates from
given function $R_\Gamma(\theta,\varphi)$.
At the point $x=y=z=0$ the function itself $\tilde f(x,y)$ and its first derivatives by construction are equal to zero.

The next terms of the expansion in its Fourier series form a quadratic form
\begin{equation}
\frac{\partial^2 \tilde f}{\partial x^2}x^2+2\frac{\partial^2 \tilde f}{\partial x\partial y}xy+
\frac{\partial^2 \tilde f}{\partial y^2}y^2.
\label{eq:17}
\end{equation}

If this quadratic form is not negative definite, then in a sufficiently small neighborhood, where the terms of the series
of the higher orders can be neglected, the function on some rays will have positive values.
This means finding the points of the region above the tangent plane, which contradicts the convexity of the domain.
For the quadratic form to be negative definite, the well-known inequalities for the coefficients must be satisfied.
For (\ref{eq:17}) they take form:
\begin{equation}
\frac{\partial^2 \tilde f}{\partial x^2}\leq 0,\quad
\frac{\partial^2 \tilde f}{\partial y^2}\leq 0,\quad
\left|\frac{\partial^2 \tilde f}{\partial x\partial y}\right|^2\leq
\frac{\partial^2 \tilde f}{\partial x^2}\frac{\partial^2 \tilde f}{\partial y^2}.
\label{eq:44.5}
\end{equation}

On the boundary $\gamma$, we denote a function similar to $\tilde f(x,y)$ by $f(x,y)$
and require the fulfillment of the inequality:
\begin{eqnarray}
\frac{\partial^2 f}{\partial x^2}+\frac{\partial^2 f}{\partial y^2}\leq\frac{2}{R},
\label{eq:44.7}
\end{eqnarray}
where $R$ is a given positive constant.
This condition is satisfied, for example, if the surface $\gamma$ bounds a convex region and $R$ is the minimum radius of curvature.

Boundedness condition for the angles between the normals to the surfaces $\Gamma$ and $\gamma$ and the radial
direction we write down in a form convenient for use.
The radial components of the unit normal vectors are considered to be bounded from below:
\begin{eqnarray}
n_{_\Gamma,r}\geq \xi_1, \quad -n_{\gamma,r}\geq \xi_1,
\label{eq:45}
\end{eqnarray}
where $\xi_1$ is a strictly positive constant.

We also require that the scalar product of the normals ${\bf n_{_\Gamma}}$ and ${\bf n_{_\gamma}}$
calculated on one ray is positive:
\begin{eqnarray}
{\bf n_{_\Gamma}}(\theta,\varphi)\cdot {\bf n_{_\gamma}}(\theta,\varphi)\geq \xi_2>0.
\label{eq:9}
\end{eqnarray}

For $\xi_1>1/\sqrt{2}$ this restriction can be omitted as an independent one.
Indeed, let us separate the radial component of these vectors. By virtue of (\ref{eq:45}), the other components modulo
do not exceed $\sqrt{1-\xi_1^2}\geq 0$. Therefore, all scalar product $>2\xi_1^2-1>0$,
and the last expression can simply be denoted by $\xi_2$.

We also require that the differences between both surfaces and two spheres with a common center and not too different radii,
is limited. Namely, let the constants introduced above satisfy the inequality:
\begin{eqnarray}
\frac{2R_2^2\delta R}{\xi_1\xi_2^2R_1^2R}<1.
\label{eq:10}
\end{eqnarray}

In mathematical modeling of the D-layer of the ionosphere, this condition is met with a large margin, the fraction is less than $0.02$.
If the boundaries $\gamma$ and $\Gamma$ are the concentric spheres with radii $R_1$ and $R_3$,
then $\xi_1=\xi_2=1$, $R=R_1$, and (\ref{eq:10}) becomes the condition $R_3/R_1<1.5$.

\section{Estimate of the $L_2$ norm of the trace of a vector function on the inner part of the boundary}
In this section, we prove the first of the inequalities
\begin{eqnarray}
\oint_\gamma |{\bf P}|^2\,d\gamma\leq C_3\int_\Omega \left|\mbox{grad}\,{\bf P}\right|^2\,d\Omega, \quad
\oint_\Gamma |{\bf P}|^2\,d\Gamma\leq C_4\int_\Omega \left|\mbox{grad}\,{\bf P}\right|^2\,d\Omega,
\label{eq:1}
\end{eqnarray}
where the constants $C_3,C_4$ are determined only by the shape of the region,
and in the next section - the second one.

These inequalities mean the estimates of the  $L_2$ norms of the traces of the vector function ${\bf P}$ on two sections of the boundary.
Without concrete values of the constants, they follow from the theorems
of embedding of the space $W^{(1)}_2(\Omega)$ to the space $L_2(\Gamma)$ \cite{Sobolev},
if we additionally require that the values of the Cartesian components ${\bf P}$ averaged over $\Omega$ are equal to zero.

Consider the ray segment $\theta=const$, $\varphi=const$ from $\gamma$ to $\Gamma$.
We fix the normal ${\bf n}_{_\Gamma}$ at the point of intersection of the ray with the surface $\Gamma$ and denote
\begin{eqnarray}
\alpha(r)={\bf n}_{_\Gamma}\cdot {\bf P}.
\label{eq:45.1}
\end{eqnarray}

Since the first factor does not vary along the ray,
\begin{eqnarray}
\frac{\partial\alpha(r)}{\partial r}={\bf n}_{_\Gamma}\cdot \frac{\partial}{\partial r}{\bf P}.
\label{eq:45.2}
\end{eqnarray}

Taking into account that $\alpha$ equals to zero at $\Gamma$ by the second condition (\ref{eq:11}),
by the Newton-Leibniz formula we obtain at a point on $\gamma$:
\begin{eqnarray}
\alpha=\int_{r=R_\Gamma(\theta,\varphi)}^{r=R_\gamma(\theta,\varphi)}\frac{\partial\alpha(r)}{\partial r}\,dr.
\nonumber
\end{eqnarray}

We use the Cauchy-Bunyakovsky inequality taking into account the constraint (\ref{eq:5}), before the last one:
\begin{eqnarray}
\alpha^2\leq\delta R\int_{r=R_\gamma(\theta,\varphi)}^{r=R_\Gamma(\theta,\varphi)}
\left|\frac{\partial\alpha(r)}{\partial r}\right|^2\,dr.
\label{eq:46}
\end{eqnarray}

Since ${\bf n}_{_\Gamma}$ is a unit vector
\begin{eqnarray}
\left|\frac{\partial\alpha(r)}{\partial r}\right|^2\leq
\left|\frac{\partial}{\partial r}{\bf P}\right|^2 \leq
\left|\mbox{grad}\,{\bf P}\right|^2.
\nonumber
\end{eqnarray}

This allows to continue the evaluation (\ref{eq:46}):
\begin{eqnarray}
\alpha^2\leq\delta R\int_{r=R_\gamma(\theta,\varphi)}^{r=R_\Gamma(\theta,\varphi)}
\left|\mbox{grad}\,{\bf P}\right|^2\,dr.
\nonumber
\end{eqnarray}

At $\gamma$, by the last condition (\ref{eq:11}), the tangential components of the vector ${\bf P}$ are equal to zero,
therefore ${\bf P}=|{\bf P}|{\bf n_{_\gamma}}$. Consequently
$$
\alpha=|{\bf P}|{\bf n_{_\gamma}}\cdot {\bf n_{_\Gamma}},
$$
and due to (\ref{eq:9})
$$
|{\bf P}|\leq\alpha/\xi_2.
$$

The normal component of the vector also satisfies this inequality, especially since the tangential ones are equal to zero.
We strengthen the inequality by introducing a factor that is greater than $1$ due to (\ref{eq:5}):
\begin{eqnarray}
P_n^2\leq\frac{\delta R}{\xi_2^2}\int_{r=R_\gamma(\theta,\varphi)}^{r=R_\Gamma(\theta,\varphi)}
\left|\mbox{grad}\,{\bf P}\right|^2\,\frac{r^2}{R_1^2}\,dr.
\label{eq:49}
\end{eqnarray}

We integrate this inequality over $\gamma$ and reduce the integration over
surface to integration over the spherical angles using the equality:
\begin{eqnarray}
\oint_\gamma P_n^2 \ d\gamma =\oint_\gamma P_n^2\,
\frac{d\gamma}{R^2_\gamma(\theta,\varphi)\sin{\theta}\,d\theta d\varphi} R^2_\gamma(\theta,\varphi)\,\sin{\theta}\,d\theta d\varphi,
\label{eq:50}
\end{eqnarray}
where the fraction equals to the ratio of the area at $\gamma$ to the area of the projection
of this surface element onto the sphere. By virtue of (\ref{eq:45}) it does not exceed $1/\xi_1$.

Substituting (\ref{eq:49}) in (\ref{eq:50}) and replacing $R_\gamma(\theta,\varphi)$ with its maximum value $R_2$,
we obtain an integral over the domain $\Omega$:

\begin{eqnarray}
\oint_\gamma P_n^2 \ d\gamma\leq \frac{R_2^2\delta R}{R_1^2\xi_1\xi_2^2}
\int_\Omega \left|\mbox{grad}\,{\bf P}\right|^2\,r^2\,\sin{\theta}\,dr\,d\theta\,d\varphi.
\nonumber
\end{eqnarray}

Since the tangential components of the vector ${\bf P}$ at $\gamma$ are equal to zero,
this proves the first inequality (\ref{eq:1}) with the constant
\begin{eqnarray}
C_3=\frac{R_2^2\delta R}{R_1^2\xi_1\xi_2^2}.
\label{eq:52}
\end{eqnarray}

\section{Estimate of the $L_2$ norm of the trace of a vector function on the outer part of the boundary}
Consider the ray segment $\theta=const$, $\varphi=const$ from $\gamma$ to $\Gamma$.
At the point of intersection of the ray with $\gamma$, we introduce Cartesian coordinates with the $z$ axis along the outward normal to $\gamma$.
We denote the projections of the vector ${\bf P}$ on the axis $x$, $y$, $z$ as $P_x$, $P_y$, $P_z$, respectively.

Taking into account the equality to zero of $P_x$, $P_y$ on $\gamma$ due to the first condition
(\ref{eq:11}), using the Newton-Leibniz formula, we obtain at a point on
$\Gamma$:
\begin{eqnarray}
P_x=\int_{r=R_\gamma(\theta,\varphi)}^{r=R_\Gamma(\theta,\varphi)}\frac{\partial P_x(r)}{\partial r}\,dr,  \quad
P_y=\int_{r=R_\gamma(\theta,\varphi)}^{r=R_\Gamma(\theta,\varphi)}\frac{\partial P_y(r)}{\partial r}\,dr.
\nonumber
\end{eqnarray}

Applying the Cauchy-Bunyakovsky inequality and taking into account before the last constraint (\ref{eq:5}), we obtain:
\begin{eqnarray}
P_x^2\leq\delta R\int_{r=R_\gamma(\theta,\varphi)}^{r=R_\Gamma(\theta,\varphi)}
\left|\frac{\partial P_x(r)}{\partial r}\right|^2\,dr, \quad
P_y^2\leq\delta R\int_{r=R_\gamma(\theta,\varphi)}^{r=R_\Gamma(\theta,\varphi)}
\left|\frac{\partial P_y(r)}{\partial r}\right|^2\,dr.
\label{eq:56}
\end{eqnarray}

Since $\partial/\partial r$ is one of the components of the vector ${\mbox{grad}}$
\begin{eqnarray}
\left|\frac{\partial P_x(r)}{\partial r}\right|^2\leq
\left|\mbox{grad}\,P_x\right|^2, \quad
\left|\frac{\partial P_y(r)}{\partial r}\right|^2\leq
\left|\mbox{grad}\,P_y\right|^2.
\nonumber
\end{eqnarray}

This allows to continue the evaluation (\ref{eq:56}).
Let's strengthen inequalities by introducing a factor that is greater than $1$ due to (\ref{eq:5}):
\begin{eqnarray}
P_x^2\leq\delta R\int_{r=R_\gamma(\theta,\varphi)}^{r=R_\Gamma(\theta,\varphi)}
\left|\mbox{grad}\,P_x\right|^2\,\frac{r^2}{R_1^2}\,dr, \quad
P_y^2\leq\delta R\int_{r=R_\gamma(\theta,\varphi)}^{r=R_\Gamma(\theta,\varphi)}
\left|\mbox{grad}\,P_y\right|^2\,\frac{r^2}{R_1^2}\,dr.
\label{eq:58.1}
\end{eqnarray}

We fix the normal ${\bf n}$ at the point of intersection of the ray with the boundary $\Gamma$.
Due to the first condition (\ref{eq:11}), the normal component of the vector ${\bf P}$ is equal to zero at $\Gamma$.
Therefore, at a point at $\Gamma$:
\begin{eqnarray}
\left({\bf P} \cdot {\bf n}\right) = P_x n_x + P_y n_y + P_z n_z = 0.
\label{eq:58.2}
\end{eqnarray}

Hence,
\begin{eqnarray}
P_z^2=\left((P_x n_x + P_y n_y)/n_z\right)^2.
\label{eq:58.21}
\end{eqnarray}

Applying the Cauchy-Bunyakovsky inequality for the numerator, we obtain
\begin{eqnarray}
\left(P_x n_x + P_y n_y\right)^2\leq \left(P_x^2 + P_y^2\right)\left(n_x^2 + n_y^2\right).
\label{eq:58.22}
\end{eqnarray}

Due to (\ref{eq:9}) $n_z\geq\xi_2$. So
\begin{eqnarray}
n_x^2 + n_y^2 = 1-n_z^2 \leq 1-\xi_2^2.
\label{eq:58.24}
\end{eqnarray}

The (\ref{eq:58.1}, \ref{eq:58.21}, \ref{eq:58.22}, \ref{eq:58.24}) result into
\begin{eqnarray}
P_z^2\leq\frac{\delta R{(1-\xi_2^2)}}{\xi_2^2}
\int_{r=R_\gamma(\theta,\varphi)}^{r=R_\Gamma(\theta,\varphi)}
\left(\left|\mbox{grad}\,P_x\right|^2 + \left|\mbox{grad}\,P_y\right|^2\right)\,\frac{r^2}{R_1^2}\,dr.
\label{eq:58.3}
\end{eqnarray}

Since all Cartesian components (\ref{eq:58.1}, \ref{eq:58.3}) are estimated, the vector {\bf P} satisfies the estimate
\begin{eqnarray}
|{\bf P}|^2\leq\frac{\delta R}{\xi_2^2R_1^2}
\int_{r=R_\gamma(\theta,\varphi)}^{r=R_\Gamma(\theta,\varphi)}
\left|\mbox{grad}\,{\bf P}\right|^2r^2\,dr.
\label{eq:59}
\end{eqnarray}

We integrate the inequality (\ref{eq:59}) over $\Gamma$ and reduce the integration over
the surface to integration over the spherical angles using the equality:
\begin{eqnarray}
\oint_\Gamma |{\bf P}|^2 \ d\Gamma =\oint |{\bf P}|^2\,
\frac{d\Gamma}{R^2_\Gamma(\theta,\varphi)\sin{\theta}\,d\theta d\varphi} R^2_\Gamma(\theta,\varphi)\,\sin{\theta}\,d\theta\,d\varphi,
\label{eq:60}
\end{eqnarray}
where the fraction equals to the ratio of the area at $\Gamma$ to the area of the projection
of this surface element onto the sphere. By virtue of (\ref{eq:45}) it does not exceed $1/\xi_1$.

Substituting (\ref{eq:59}) in (\ref{eq:60}), and replacing $R_\Gamma(\theta,\varphi)$ with its maximum value $R_3$,
we obtain in the right-hand side the integral over the domain $\Omega$:
\begin{eqnarray}
\oint_\Gamma |{\bf P}|^2 \ d\Gamma\leq \frac{\delta R R_3^2}{R_1^2\xi_1\xi_2^2}
\int_\Omega \left|\mbox{grad}\,{\bf P}\right|^2\,r^2\,\sin{\theta}\,dr\,d\theta\,d\varphi.
\nonumber
\end{eqnarray}

Thus, we have proved the second inequality (\ref{eq:1}) with the constant
\begin{eqnarray}
C_4=\frac{\delta R R_3^2}{R_1^2\xi_1\xi_2^2}.
\nonumber
\end{eqnarray}

\section{Estimate of the $L_2$ norm of a vector function}
In this section, we prove the inequality (\ref{eq:3}).

Consider the ray segment $\theta=const$, $\varphi=const$ from $\gamma$ to $\Gamma$, and divide it
into two parts with some point $A$ inside the domain $\Omega$.
Let us fix the normal ${\bf n}$ at the point of intersection of the ray with the surface $\Gamma$.
Arguments similar to those used in Section 2 give an estimate for $\alpha$ at the point $A$
\begin{eqnarray}
\alpha^2\leq\delta R\int_{r=R_A(\theta,\varphi)}^{r=R_\Gamma(\theta,\varphi)}
\left|\mbox{grad}\,{\bf P}\right|^2\,\frac{r^2}{R_1^2}\,dr.
\label{eq:63}
\end{eqnarray}

Now consider a segment of the same ray from the opposite boundary $\gamma$ to the same point $A$.
The point of intersection of the ray with $\gamma$ is taken as the origin of the Cartesian coordinates. The $z$ axis is directed along the outward normal to $\gamma$.
Reasoning similar to those used in Section 3 gives an estimate for $P_x$, $P_y$ at the point $A$
\begin{eqnarray}
P_x^2\leq\delta R\int_{r=R_\gamma(\theta,\varphi)}^{r=R_A(\theta,\varphi)}
\left|\mbox{grad}\,P_x\right|^2\,\frac{r^2}{R_1^2}\,dr, \quad
P_y^2\leq\delta R\int_{r=R_\gamma(\theta,\varphi)}^{r=R_A(\theta,\varphi)}
\left|\mbox{grad}\,P_y\right|^2\,\frac{r^2}{R_1^2}\,dr.
\label{eq:63.1}
\end{eqnarray}

Since the scalar product does not exceed the product of the modulus of the vectors, $|{\bf n}|=1$, and the integration in
(\ref{eq:63}, \ref{eq:63.1}) is made over two different parts of the segment under consideration, at the point $A$
\begin{eqnarray}
\left|{\bf P} \cdot {\bf n}\right| = \left|P_x n_x + P_y n_y + P_z n_z \right| \leq
 \sqrt{\delta R\int_{r=R_\gamma(\theta,\varphi)}^{r=R_\Gamma(\theta,\varphi)} \left|\mbox{grad}\,{\bf P}\right|^2\,\frac{r^2}{R_1^2}\,dr}.
\nonumber
\end{eqnarray}

This implies
\begin{eqnarray}
\left|P_z \right| \leq \frac{1}{n_z} \left(\sqrt{\delta R\int_{r=R_\gamma(\theta,\varphi)}^{r=R_\Gamma(\theta,\varphi)}
\left|\mbox{grad}\,{\bf P}\right|^2\,\frac{r^2}{R_1^2}\,dr} + \left|P_x n_x + P_y n_y \right|\right).
\label{eq:64.2}
\end{eqnarray}

Since the scalar product does not exceed the product of the absolute values of the vectors, $|{\bf n}|=1$,
and due to (\ref{eq:45}) $n_z\geq\xi_2$,
\begin{eqnarray}
\left|P_x n_x + P_y n_y\right|\leq \sqrt{1-\xi_2^2}\sqrt{P_x^2 + P_y^2}.
\nonumber
\end{eqnarray}

Taking this inequality into account, we square both sides (\ref{eq:64.2}) and get
\begin{eqnarray}
P_z^2 \leq \frac{1}{\xi_2^2} \left(\sqrt{\delta R\int_{r=R_\gamma(\theta,\varphi)}^{r=R_\Gamma(\theta,\varphi)}
\left|\mbox{grad}\,{\bf P}\right|^2\,\frac{r^2}{R_1^2}\,dr} + \sqrt{1-\xi_2^2}\sqrt{P_x^2 + P_y^2}\right)^2.
\label{eq:64.4}
\end{eqnarray}

For the right-hand side (\ref{eq:64.4}), we apply the inequality $(a+b)^2\leq 2(a^2+b^2)$, which is valid for any real numbers
\begin{eqnarray}
P_z^2 \leq \frac{2}{\xi_2^2} \left(\delta R\int_{r=R_\gamma(\theta,\varphi)}^{r=R_\Gamma(\theta,\varphi)}
\left|\mbox{grad}\,{\bf P}\right|^2\,\frac{r^2}{R_1^2}\,dr + (1-\xi_2^2)(P_x^2 + P_y^2)\right).
\label{eq:64.5}
\end{eqnarray}

Due to (\ref{eq:45}) $1-\xi_2^2\leq 1$. Therefore, using (\ref{eq:63.1}) for (\ref{eq:64.5}), we get
\begin{eqnarray}
P_z^2 \leq \frac{2\delta R}{\xi_2^2R_1^2}
\int_{r=R_\gamma(\theta,\varphi)}^{r=R_\Gamma(\theta,\varphi)}
\left|\mbox{grad}\,{\bf P}\right|^2r^2\,dr,
\label{eq:65}
\end{eqnarray}

Using (\ref{eq:63.1}, \ref{eq:65}), estimate the whole vector ${\bf P}$ at the point $A$
\begin{eqnarray}
|{\bf P}|^2\leq\frac{2\delta R}{\xi_2^2R_1^2} \int_{r=R_\gamma(\theta,\varphi)}^{r=R_\Gamma(\theta,\varphi)}
\left|\mbox{grad}\,{\bf P}\right|^2r^2\,dr + \frac{\delta R}{R_1^2} \int_{r=R_\gamma(\theta,\varphi)}^{r=R_A(\theta,\varphi)}
\left|\mbox{grad}\,{\bf P}\right|^2r^2\,dr.
\label{eq:65.2}
\end{eqnarray}

Due to (\ref{eq:45}) $2+\xi_2^2 \leq 3$ and if we extend the limits of integration of the positive function in the last integral (\ref{eq:65.2}),
then we get
\begin{eqnarray}
|{\bf P}|^2\leq\frac{3\delta R}{\xi_2^2R_1^2} \int_{r=R_\gamma(\theta,\varphi)}^{r=R_\Gamma(\theta,\varphi)}
\left|\mbox{grad}\,{\bf P}\right|^2r^2\,dr.
\nonumber
\end{eqnarray}

We multiply this inequality by $ r ^ 2 $ and integrate it over $ r $:
\begin{eqnarray}
\int_{r=R_\gamma(\theta,\varphi)}^{r=R_\Gamma(\theta,\varphi)} |{\bf P}|^2r^2 dr \leq
\frac{3\delta R}{\xi_2^2R_1^2}\int_{\tilde{r}=R_\gamma(\theta,\varphi)}^{\tilde{r}=R_\Gamma(\theta,\varphi)}
\left|\mbox{grad}\,{\bf P}\right|^2 \tilde{r}^2 d\tilde{r} \int_{r=R_\gamma(\theta,\varphi)}^{r=R_\Gamma(\theta,\varphi)}r^2 dr.
\nonumber
\end{eqnarray}

Here the right-hand side (\ref{eq:65.2}) is immediately removed from the integral, since it does not depend on $r$.
Due to (\ref{eq:5}) $r^2\leq R_3^2$, and the integration interval does not exceed $\delta R$. Consequently
\begin{eqnarray}
\int_{r=R_\gamma(\theta,\varphi)}^{r=R_\Gamma(\theta,\varphi)} |{\bf P}|^2 r^2 dr \leq
3\left(\frac{\delta R R_3}{\xi_2 R_1}\right)^2 \int_{r=R_\gamma(\theta,\varphi)}^{r=R_\Gamma(\theta,\varphi)}
\left|\mbox{grad}\,{\bf P}\right|^2 r^2 dr.
\nonumber
\end{eqnarray}

Multiplying both sides of this inequality by $\sin\theta$, and integrating over $\theta$ and $\varphi$,
we prove the inequality (\ref{eq:3}) with the constant
\begin{eqnarray}
C_1=3\left(\delta R R_3/(\xi_2 R_1)\right)^2.
\nonumber
\end{eqnarray}

\section{Estimate of the $L_2$ norm of the gradient of a vector function}
In the paper \cite{Bykhovsky} an identity is given, which is easy to verify for multiply connected domains:
\begin{eqnarray}
\int ((\mbox{rot}\ {\bf P})^2 + (\mbox{div}\ {\bf P})^2) \ d\Omega=
\int (\mbox{grad}\ {\bf P})^2 \ d\Omega+\nonumber\\
+\oint_\Gamma ((\mbox{div}\ {\bf P})P_n-({\bf P}\mbox{grad})P_n) \ d\Gamma
+\oint_\gamma ((\mbox{div}\ {\bf P})P_n-({\bf P}\mbox{grad})P_n) \ d\gamma.
\label{eq:44.2}
\end{eqnarray}

Using the expression for the surface equation in the form (\ref{eq:16}), the integrand in the integral over the boundary $\Gamma$
with simple but cumbersome calculations can be converted to the form
\begin{eqnarray}
-(\frac{\partial^2 \tilde f}{\partial x^2}+\frac{\partial^2 \tilde f}{\partial y^2})P_z^2 -\frac{\partial^2 \tilde f}{\partial
x^2}P_x^2 -2\frac{\partial^2 \tilde f}{\partial x\partial y}P_xP_y -\frac{\partial^2 \tilde f}{\partial y^2}P_y^2
+(\frac{\partial P_x}{\partial x}+\frac{\partial P_y}{\partial y})P_z
-\frac{\partial P_z}{\partial x}P_x-\frac{\partial P_z}{\partial y}P_y.
\label{eq:44.3}
\end{eqnarray}

Because the normal compponent of the ${\bf P}$ is zero (\ref{eq:11}), $P_z=0$, $\partial P_z/\partial x=0$, $\partial
P_z/\partial y=0$, and in (\ref{eq:44.3}) on $\Gamma$ only the quadratic form remains
\begin{equation}
-\frac{\partial^2 \tilde f}{\partial x^2}P_x^2 -2\frac{\partial^2
\tilde f}{\partial x\partial y}P_xP_y -\frac{\partial^2 \tilde f}{\partial y^2}P_y^2.
\label{eq:44.4}
\end{equation}

It only differs in the sign of the coefficients from (\ref{eq:17}), and therefore is positive definite.
Therefore, the integral over $\Gamma$ in (\ref{eq:44.2}) is non-negative.
On $\gamma$, the tangential components of ${\bf P}$ are equal to zero
(\ref{eq:11}). Therefore $P_x=P_y=0$, $\partial P_x/\partial x=0$,
$\partial P_y/\partial y=0$, and in a similar (\ref{eq:44.3}) expression, only
\begin{eqnarray}
-(\frac{\partial^2 f}{\partial x^2}+\frac{\partial^2 f}{\partial y^2})P_z^2.
\label{eq:44.6}
\end{eqnarray}

Using the resulting expression for the integrand (\ref{eq:44.6}) and the constraint (\ref{eq:44.7}),
estimate the integral over $\gamma$ in (\ref{eq:44.2}) from above:
\begin{eqnarray}
\left|\oint_\gamma ((\mbox{div}\ {\bf P})P_n-({\bf P}\mbox{grad})P_n)\,d\gamma\right|\leq
\frac{2}{R}\oint_\gamma P_n^2\,d\gamma.
\label{eq:44.8}
\end{eqnarray}

Taking into account the first inequality (\ref{eq:1}), we obtain the estimate
\begin{eqnarray}
\left|\oint_\gamma ((\mbox{div}\ {\bf P})P_n-({\bf P}\mbox{grad})P_n)\,d\gamma\right|\leq
\frac{2C_3}{R}\int_\Omega \left|\mbox{grad}\,{\bf P}\right|^2\,d\Omega.
\label{eq:44.9}
\end{eqnarray}

Since the integral over $\Gamma$, as already shown, is non-negative, and the integral over $\gamma$ is estimated from above in the
inequality (\ref{eq:44.9}), taking into account the expression $C_3$ (\ref{eq:52}), we obtain
\begin{eqnarray}
\int ((\mbox{rot}\ {\bf P})^2 + (\mbox{div}\ {\bf P})^2)\,d\Omega\geq
(1-\frac{2R_2^2\delta R}{\xi_1\xi_2^2R_1^2R})\int (\mbox{grad}\ {\bf P})^2\,d\Omega.
\label{eq:44.10}
\end{eqnarray}

For this estimate to be meaningful, the factor in front of the integral must be positive, i.e. the fraction must be less than $1$.
For this, the condition (\ref{eq:10}) was imposed.

\section{Conclusion}
Thus, for vector functions ${\bf P}$ satisfying the mixed boundary conditions of the form (\ref{eq:11}) in a multiply connected domain,
inequalities are proved which are necessary in the study of the operators of elliptic boundary value problems.
The inequality (\ref{eq:3}) estimates from above the $L_2$ norm of a vector function in terms of $L_2$ the norm of its gradient.
The inequality (\ref{eq:4}) estimates from above the $L_2$ norm of the gradient of a vector function through the sum of the $L_2$ norms of its
rotor and divergence.
The inequalities (\ref{eq:1}) for the $L_2$ norms of the traces of such vector functions on both parts of the boundary are also obtained.

Note that the condition of convexity of the outer part of the boundary $\Gamma$ and the constraints (\ref{eq:44.5}) can be weakened,
only slightly complicating the proof.
Then the integral over $\Gamma$ in (\ref{eq:44.2}) will no longer be non-negative, but it can be estimated in the same way as the integral over $\gamma$.
This will lead to the appearance of an additional negative term in the factor before the integral (\ref{eq:44.10}),
and, therefore, will require a stronger constraint for the constants (\ref{eq:10}).

It should also be said that the used restrictions on the shape of the domain are associated only with the possibility
to obtain simple proofs, and the inequalities themselves, apparently, are valid for the same multiply connected domains of general form,
for which these inequalities were proved in the work \cite{Bykhovsky_Smirnov} under one of the conditions (\ref{eq:11})
on the whole boundary.

\subsection*{Acknowledgments}
The research is supported by Russian Foundation for Basic Research (project 18-05-00195).

\end{document}